\documentstyle[prd,aps,eqsecnum,psfig]{revtex}

\newcommand{\be}{\begin{equation}}
\newcommand{\ee}{\end{equation}}
\newcommand{\bea}{\begin{eqnarray}}
\newcommand{\eea}{\end{eqnarray}}
\newcommand{\bm}{\bibitem}
\newcommand{\om}{\omega}
\newcommand{\al}{\alpha}
\newcommand{\bt}{\beta}
\newcommand{\sg}{\sigma}
\newcommand{\Sg}{\Sigma}
\newcommand{\de}{\delta}

\newcommand{\gm}{\gamma}
\newcommand{\Gm}{\Gamma}
\newcommand{\ep}{\epsilon}

\newcommand{\ps}{p \!\!\! /}
\newcommand{\call}{{\cal L}_I}
\newcommand{\vk}{\vec{k}}
\newcommand{\vp}{\vec{p}}
\newcommand{\la}{\langle}
\newcommand{\ra}{\rangle}

\newcommand{\ipx}{ip\cdot x}
\newcommand{\ad}{a^{\dag}}
\newcommand{\spn}{\Sigma_f^{(0)} (p)}
\newcommand{\stp}{{\Sigma}_f (p,u)}
\newcommand{\stu}{\tilde{\Sigma}_f (p,u)}
\newcommand{\bu}{\bullet}

\begin{document}

\title{Scattering amplitude and shift in self-energy in medium}

\author{S. Mallik\footnote{Permanent address : Saha Institute of Nuclear
Physics, 1/AF, Bidhannagar, Calcutta-700064, India}}
\address{Institute for Theoretical Physics, University of Bern,
Sidlerstrasse 5, CH-3012 Bern, Switzerland}

\maketitle

\begin{abstract}
Two simple proofs are presented for the first order virial expansion of the
self-energy of a particle moving through a medium,
characterised by temperature and/or chemical potential(s). One is based
on the virial expansion of the self-energy operator itself, while the other is
on the analysis of its Feynman diagrams in configuration space.
\end{abstract}



\renewcommand{\thefootnote}{\arabic{footnote}}
\setcounter{footnote}{0}


\section{Introduction}
\label{sec:Introduction}

More than a decade ago, Leutwyler and Smilga \cite{Leutwyler}
considered the problem of mass-shift and damping rate of a nucleon
propagating through a heat bath. As one of the methods to deal with the
problem, they wrote (the first term of)
the virial expansion of the nucleon self-energy, which relates it to the
pion-nucleon scattering amplitude in the forward direction. At low pion
density the first order formula yields the dominating contribution.
The advantage of such a formula is that one can use the experimental
data to compute the shift in self-energy without relying on any theoretical
structure. Further the formula is simple enough to suggest a generalisation
to other hadrons in different media. Accordingly a number of authors
have used these relations to explore the properties of hadrons in
such media \cite{Shuryak,Eletsky1,Eletsky2}.

In the context of quantum electrodynamics, Barton \cite{Barton} took a
physical approach to derive such formulae, following an earlier
suggestion by Feynman \cite{Feynman}. The refractive index of a medium for a
beam of particles of a definite frequency differs from unity due to the
scattering of the incident beam by particles of the medium.
The change in frequency (energy) of the
particles may be related to this change in refractive index. In this way
Feynman related the self-energy of the electron to its forward scattering
amplitude with the virtual quanta
in vacuum. Barton extended these considerations to the black body
radiation.

Recently the complete expansion of the self-energy in the medium in powers
of the distribution function has been proven in perturbation
theory \cite{Jeon}.
In effect, it is a reordering of the contributions in perurbation theory,
which is an expansion in powers of coupling contant, to one in powers of the
distribution function. The analysis is carried
out in imaginary time formulation of field theory in the medium.

Here we present two simple proofs of the first term in the expansion of the
self-energy. One is a generalisation of the method used in
Ref.\cite{Klingl} and is based directly on the virial expansion of the
ensemble average of any operator. The other is in perturbation
theory, based on Feynman diagrams in configuration space.

We shall work here in the real time formulation of field theory at finite
temperature and density. Secs. II and III describe the two proofs of the
virial expansion. Sec. IV discusses the relevance of such formulae and
possibility of extending the proofs to higher order.

\section{Virial expansion}
\label{sec:virial expansion}

Following Weinberg \cite{Weinberg}, we introduce a compact notation
for different types particles and their fields.
Let $a(\vp,\sg,f)$ and $\ad (\vp,\sg,f)$ be the
destruction and creation operators for a particle of species $f$ having
momentum $\vp$ and spin projection $\sg$; then for example,
\[ |\vp,\sg,f \ra = \ad (\vp,\sg,f) |0\ra  .\]
They satisfy the commutation  (anticommutation) relations
$(\om_p=\sqrt{\vp\,^2+m^2})$,
\[ [a(\vp,\sg,f), \ad(\vp\, ',\sg',f']_{\mp}=(2\pi)^3 2\om_p
\de^3(\vp -\vp\,') \de_{\sg\sg'}\de_{ff'} . \]
The field operators are denoted by $\psi_l(x)$, where the index $l$ denotes
not only the field type but also runs over its components. Then $\psi_l(x)$
may be expanded as
\be
\psi_l(x)=\sum_\sg \int{d^3p\over {(2\pi)^3}}{1\over {2\om_p}}
\left [ u_l(\vp,\sg,f) a(\vp,\sg,f) e^{-\ipx} + v_l(\vp,\sg,f)
\ad (\vp,\sg,f^c) e^{\ipx}\right ],
\ee
$f^c$ denoting the antiparticle of the species $f$.
The coefficient functions $u_l$ and $v_l$ depend on the
spin of the
particle. We have in mind the three types of fields, namely, the scalar
field $\phi (x)$ for which $u_l=v_l=1$; the Dirac spinor field $\psi (x)$, for
which they are the normalised Dirac spinors, $\bar{u}(\vp,\sg) u(\vp,\sg)
=-\bar{v}(\vp,\sg)v(\vp,\sg)=2m $ and the vector field $A_{\mu}$, for which
they are the polarisation vectors, $\ep_{\mu} (\vp, \sg), \,
\ep_{\mu}^* (\vp, \sg)  \ep^{\mu} (\vp, \sg ') =-\de_{\sg\sg '}$.
Below we shall in most places suppress the variable $\sg$ of $u_l, \, v_l,\,
a\,$ and $\ad$. If there is an integration over a 3-momentum, a summation
over the corresponding $\sg$ as in Eq (2.1) will also be implied.

In the real time formulation of field theory in medium,
all Greens functions, in particular the
self-energy, acquire a $2\otimes 2$
matrix structure. Since we confine here only to contributions linear in the
distribution function, it will, however, suffice to consider only the
11-component of these matrices. Below we shall drop
the 11 index \footnote{ The self-energy function $\Sg $ which actually
shifts the pole
position of the propagator is related to the components in a
simple way \cite{Kobes1}.
With $\Sg_{11}$ these relations are, $Re\Sg (p)=Re\Sg _{11} (p)$ and
$Im \Sg(p) =\al Im \Sg_{11} (p)$, where $\al= tanh(\bt p_0/2)$ for bosons
and $\al=coth \{\bt(p_0 -\mu)/2\}$ for fermions ($p_o>0)$. Thus
$\Sg_{11}$ deviates from
$\Sg$ by exponential corrections at small temperatures. Further the factor
$\al$ actually cancels on rearranging the distribution functions within the
integral, at least for 1-loop contributions \cite{Mallik}.}.

Let us begin with the self-energy
$\spn$ of a particle of type $f$ and momentum $p$ in vacuum.
For our purpose, we write this amputated, two-point Green's function as
\be
-i(2\pi)^4\de^4(p'-p) \bar{u}_l (\vp, f)\spn u_l(\vp, f)
=\la 0 |a(\vp\, ', f) (S -1)\ad (\vp, f) |0 \ra,
\ee
where the $S$-operator is given by the familiar time ordered expression,
\[ S=T e^{i\int d^4x {\call} (x)} ,\]
${\call} (x)$ being any interaction Lagrangian built out of the fields
$\psi_l (x)$. Because $\spn$ is one-particle irreducible by definition, it
is understood here and below that we only retain such diagrams in
matrix elements like the one on the right hand side of Eq.(2.2).
The corresponding self-energy in the medium will be denoted by
$\stp$, where $u^{\mu}$ is the 4-velocity of the medium \footnote{ No
confusion should arise from the use of same $u$ in both $u^{\mu}$ and
$u_l (\vp, f)$.}. Actually we shall
work in the rest frame of the medium $(u^0 =1,\, \vec{u}=0 )$. Then we may
write
\be
-i(2\pi)^4\de^4(p'-p) \bar{u}_l (\vp, f)\stp u_l(\vp, f)
=\la a(\vp\, ', f) (S -1)\ad (\vp,\ f) \ra,
\ee
where $\la\cdots\ra$ denotes ensemble average: For any operator $O$,
\be
\la O\ra = Tr(\rho O), \qquad
\rho=e^{-\bt(H-\mu {\cal N})}/Tre^{-\bt(H-\mu {\cal N})}.
\ee
Here $H$ is the Hamiltonian, $\bt^{-1}$ the temperature. For illustration
we include a chemical potential $\mu$ for a fermionic species with the
corresponding number operator ${\cal N}$.

The ensemble average $\la O \ra $ admits a virial expansion
\be
\la O\ra = \la 0| O|0\ra + \sum_{f'}\int{d^3k\over {(2\pi)^3 2\om_k}}
n_{f'}(\om_k ) \la \vk, f' | O |\vk, f'\ra +\cdots,
\ee
where the sum over $f'$ runs in general over the species of particles
present in the medium.
The dots represent terms of higher order in the distribution function $n$.
The latter are given by the familiar expressions,
$n_B(\om_k)=(e^{\bt \om_k}-1)^{-1}$ for bosons and $n_F^{\pm} (\om_k)=
(e^{\bt(\om_k\mp\mu)}+1)^{-1}$ for fermions and antifermions respectively.
Applying Eq.(2.5) to the left hand side of Eq.(2.3), we get
for $\stu \equiv \stp -\spn$,
\be
-i(2\pi)^4\de^4(p'-p) \bar{u}_l (\vp, f)\stu u_l(\vp, f)
=\sum_{f'}\int{d^3k\over {(2\pi)^3 2\om_k}} n_{f'}(\om_k )
\la \vk, f'|a(\vp\, ', f) (S -1)\ad (\vp , f)|\vk, f'\ra .
\ee
The matrix element on the right in this equation is recognised to be the
amplitude $T$ for scattering of a $f$-particle of momentum $p$ with a
$f'$-particle of momentum $k$ in the forward direction,
\be
\la \vk, f'|a(\vp\, ', f) (S -1)\ad (\vp, f)|\vk, f'\ra
=i(2\pi)^4 \de^4 (\vp\, '-\vp) T_{f f'}(p,k) .
\ee
We thus get the virial expansion for the self energy to first order,
\be
-\bar{u}_l (\vp, f)\stu u_l(\vp, f)
=\sum_{f'}\int{d^3k\over {(2\pi)^3 2\om_k}} n_{f'}(\om_k) T_{f f'}(p,k),
\ee
where an average over the polarizations of $f$-species and a sum over the
polarizations of the different $f'$-species are understood.

We now apply this formula to two cases of interest. The first one is that
of a nucleon in a medium. Its complete propagator in the medium is
$i/(\ps-m-\tilde{\Sg}_f)$. For the nucleon at rest $(\vp =0)$ in the
rest frame of the medium, we can write $\tilde{\Sg}_f=a+b\gm^0$, getting
$\bar{u}(p) \tilde{\Sg}_f u (p) =2m_N (a+b)$. Also the complete propagator
simplifies to
\[ {i\over {p_0-(m_N+a+b)}}\cdot {1\over 2} (1+\gm^0) \].
Thus the new pole position is given by \cite{Leutwyler},
\be
\tilde{m}_N - {i\over 2}\Gm_N = m_N -{1\over 2m_N}\sum_{f'}
\int{d^3k\over {(2\pi)^3 2\om_k}} n(\om_k) T_{Nf'}(p,k) .
\ee

The other case we consider is that of a vector meson in the medium.
Let us rewrite Eq.(2.8) as
\be
-{1\over 3} \sum_{\sg} \ep_{\mu}^* (p,\sg)
\tilde{\Pi}^{\mu\nu} (p,u) \ep_{\nu} (p, \sg) =\sum_{f'}
\int{d^3k\over {(2\pi)^3 2\om_k}} n_{f'}(\om_k) T_{V f'}(p,k).
\ee
The free propagator for the vector meson is
\be
D^{(0)}_{\mu\nu} (p)=\left ( -g_{\mu\nu} + {p_{\mu} p_{\nu} \over {m_V^2}}
\right ) {i\over {p^2 -m_V^2}}
\ee
To sum the series of 1-particle reducible insertions of the polarisation
tensor, we have to decompose the latter in terms of kinematic covariants,
\be
\tilde{\Pi}_{\mu\nu}=P_{\mu\nu} \Pi_t + Q_{\mu\nu} \Pi_l,
\ee
which we choose as
\be
P_{\mu\nu}=-g_{\mu\nu} +{p_{\mu}p_{\nu}\over p^2} -{p^2\over \bar{p}^2}
\tilde{u}_{\mu} \tilde{u}_{\nu}, \qquad Q_{\mu\nu}={p^4\over \bar{p}^2}
\tilde{u}_{\mu}\tilde{u}_{\nu},
\ee
where $\om=u\cdot p, \, \bar{p}=\sqrt{\om^2 -p^2}$
and $\tilde{u}_{\mu} = u_{\mu} -\om p_{\mu}/ p^2$.
The covariants are free from singularities at $p^2=0$, but at $\vp =0$, there
is a constraint on the two amplitudes,
\be
\Pi_t(p_0, \vp =0)=p_0^2 \Pi_l(p_0,\vp=0).
\ee
Then the full propagator becomes
\be
D_{\mu\nu} = {i\over {p^2-m_V^2-\Pi_t}} P_{\mu\nu} +{i\over {p^2
-m_V^2(1+\Pi_l)}} {Q_{\mu\nu}\over m_V^2}
\ee
Thus in general the transverse and the longitudinal components suffer
different shifts in the position of the pole. But for the vector meson at
rest $(\vp=0)$  the two shifts coincide, because of the constraint
equation. We then get the same formula for the pole shift as Eq.(2.9) with
the subscript $N$ replaced by $V$.

\section{Perturbation expansion}
\label{sec: perturbation expansion}

We now attempt a perturbative proof of Eq.(2.8). We begin by expanding the
$S$-operator in Eq.(2.3) in the familiar perturbation series,
\be
-i(2\pi)^4\de^4(p'-p)\bar{u}_l (\vp,f)\stp u_l (\vp, f)
 =\sum_{N=1}^{\infty} {i^N\over {N!}}\int d^4x_1
\cdots d^4x_N F^{(N)} (x_1\cdots x_N),
\ee
where
\be
F^{(N)} (x_1\cdots x_N)= \la a(\vp\, ',f) T\{ {\call} (x_1)\cdots {\call} (x_N)\}
\ad(\vp,f) \ra _{\rm paired}.
\ee
The subscript 'paired' indicates
that $F^{(N)}$ is the sum of all connected terms obtained by pairing
(contracting) all the
operators in it in all possible ways. In other words, it represents the sum
of all connected Feynman diagrams in the $N$th order. In the following
we indicate a pairing by thick dots as superscript. The pairing between a
creation or an annihilation operator of a particle
with a field operator are given by,
\bea
a^{\bu} (\vp\, ',f) \bar{\psi}_l^{\bu} (x)
&=& e^{ip'\cdot x}\bar{u}_l (\vp\,' ,f) \nonumber  \\
\psi^{\bu}_l (x) {\ad}^{\bu} (\vp, f)
&=& e^{-ip\cdot x} u_l (\vp,f) .
\eea
We may choose to work out the self energy of a particle. But the medium
may contain antiparticles. So we also note the contractions,
\bea
a^{\bu} (\vp \,',f^c) \psi_l^{\bu} (x)
&=& e^{ip'\cdot x} v_l (\vp\,', f) \nonumber  \\
\bar{\psi}^{\bu}_l (x) {\ad}^{\bu} (\vp, f^c)
&=& e^{-ip\cdot x} \bar{v}_l (\vp,f).
\eea
Finally the pairing of two field operators results in the free propagator
in the medium \cite{Kobes2},
\be
\psi_l^{\bu} (x) \bar{\psi}_m^{\bu} (y)= \la T\psi_l (x) \bar{\psi}_m (y)\ra
.
\ee

A free propagator in the medium differs from that in vacuum if there are
like-particles in the medium. In that case, it has an extra term
containing the
distribution function $n$ of the particles in the medium and a mass-shell
$\de$-function. Thus, isolating this term amounts to putting the
internal line on mass-shell, i.e. opening the propagator into two external
lines. When the $\de$-function is integrated out, this $n$-dependent piece
in the propagator, to be denoted by a subscript $n$, becomes,
\be
\la T\psi_l (x) \bar{\psi}_m (y)\ra_n
=\int{d^3k\over {(2\pi)^3 2\om_k}}\left \{ \pm n^+_f(\om_k) e^{-ik\cdot
(x-y)} u_l(\vk,f)\bar{u}_m(\vk,f) + n_f^-(\om_k) e^{ik\cdot (x-y)}
v_l(\vk,f)\bar{v}_m(\vk,f) \right \} ,
\ee
where + (--) sign before the first term holds if the species $f$ is a
boson (fermion). The distribution functions $n^+$ and $n^-$ coincide if
there is no chemical potential.

Let us assume, for simplicity, that only one species of particles
has its free propagator altered in the medium by the additional term (3.6).
We single out this field
from the compact notation $\psi_l (x)$ and call it $\chi_l(x)$
and denote its species by $f'$ \footnote{ For self-conjugate species like
$\pi^0$, the two terms in Eq.(3.6) contribute to the same amplitude, the
direct and crossed processes being identical.}. We now wish
to collect all the linear contributions in $n$ by opening each of the
$\chi$-propagators in turn in each of the Feynman diagrams for the self
energy in the $N$ th order of its perturbation expansion (Fig. 1). To
this end we consider the sum of Feynman diagrams, to be denoted by
$F^{(N)}_{ij}(x_1,\cdots, x_N)$, containing a $\chi$-propagator between
any two vertices, say at $x_i$ and $x_j$,
\be
F^{(N)}_{ij}(x_1,\cdots,x_N)
=\la a(\vp\,' f) T\{{\call}(x_1)\cdots {\call}^{\bullet} (x_i)
\cdots{\call}^{\bullet} (x_j)\cdots{\call} (x_N)\} {\ad} (\vp, f) \ra
_{\rm paired} ,
\ee
where we explicitly indicate the pairing of the two fields at $x_i$
and $x_j$.

Consider first the indicated pairing in Eq.(3.7) before any other pairings are
carried out, ie, keeping all other operators in their respective positions
in the $T$-product. By a certain number of interchanges of the field
operators we bring them together to form
$\la \chi_l(x_i)\bar{\chi}_m(x_j) \ra$. Then we extract its $n$-dependent
contribution given by Eq.(3.6) \footnote{If $f'$ is a fermionic species,
there may be two such propagators between $x_i$ and $x_j$.
We, of course, have to extract this contribution from both of them.}.
Since we restrict to linear contributions in
$n$, we set all other in-medium propagators to their vacuum values.

In order to interpret the resulting expression in terms of a two-body
scattering amplitude, we use (3.3) to rewrite the coefficient functions
for $f'$ species in
$\la \chi_l(x_i)\bar{\chi}_m(x_j) \ra$ as pairings,
\be
e^{-ik\cdot (x_i-x_j)}u_l(\vk,f') \bar{u}_m (\vk,f')
=\chi_l^{\bu} (x_i){\ad}^{\bu}(\vk, f') a^{\bu\bu} (\vk, f')
{\bar{\chi}}_m^{\bu\bu} (x_j).
\ee
For the antiparticle species $f'^c$, we use Eq.(3.4) to write an analogous
equation,
\be
e^{ik\cdot (x_i-x_j)} v_l (\vk,f') \bar{v}_m (\vk,f')
=a^{\bu} (\vk,f'^c)\chi_l^{\bu} (x_i)
{\chi}_m^{\bu\bu} (x_j) {\ad}^{\bu\bu} (\vk,f'^c).
\ee
Now the $\chi$ fields can be brought back to their original positions
forming again the complete vertices ${\call} (x_i)$ and ${\call} (x_j)$.
We also bring the creation and the annihilation operators respectively to
the right and the left of the $T$-product getting
\bea
& &F^{(N)}_{ij,n} (x_1,\cdots,x_N) \nonumber \\
& & = \int {d^3k\over {(2\pi)^3 2\om_k}}n(\om_k)
\la 0|a(\vp\,',f) a^{\bu\bu}(\vk,f') T\{{\call}(x_1)\cdots
{\call}^{\bu}(x_i)\cdots
{\call}^{\bu \bu}(x_j)\cdots {\call}(x_N)
{\ad}^{\bu}(\vk,f'){\ad}(\vp,f)|0 \ra _{\rm paired} ,
\eea
where the subscript $n$ denotes the $n$-dependent contribution from the
$\chi$-propagator connecting $x_i$ and $x_j$.

It remains to show that we do not get
any additional sign, if the $\chi$ field
is fermionic. First note that bringing the fields $\chi (x_i)$ and
$\bar{\chi}(x_j)$ together to form the propagator
and then putting them back to their old positions can
be effected by the same set of field interchanges , one in reverse order of
the other. Thus we do not encounter any extra minus sign here. Also the
interaction Lagrangians being quadratic in fermionic $\chi$, the operators
$a(\vk, f')$ and $\ad (\vk, f')$ do not produce any minus sign while
moving through them. However, in the left hand side of
Eq (3.8) an initial interchange of $a$ and
$\ad$ is needed which produces a minus sign to cancel the minus sign in
front of the first term in Eq.(3.6).
Thus Eq.(3.10) remains valid for both fermionic and bosonic operators.

Eq.(3.10) gives the linear contribution from a definite $\chi$-propagator,
namely the one between $x_i$ and $x_j$. To get the total contribution from
all the $\chi$-propagators in the diagrams, we must allow
$a(\vk, f')$ and $\ad (\vk, f')$ to be paired with $\chi$ fields
at all vertices
in all possible ways, not just with $\chi (x_i)$ and $\bar{\chi} (x_j)$.
Thus we get
\be
F^{(N)}_{n} (x_1,\cdots,x_N)=\int{d^3k\over {(2\pi)^3 2\om_q}} n(\om_q)
\la 0|a(\vp\,',f) a(\vk,f') T\{ {\call} (x_1)\cdots {\call} (x_N)\}
\ad (\vk,f') \ad (\vp,f)|0\ra _{\rm paired}.
\ee
This matrix element is just the sum of all Feynman diagrams in
coordinate space in the $N$-th order of perturbation expansion
for the scattering amplitude $T_{ff'}$ introduced previously by Eq.(2.7).
We thus prove Eq.(2.8) in an arbitrary order of pertubation theory.

\begin{figure}[!h]
\centerline{\psfig{figure=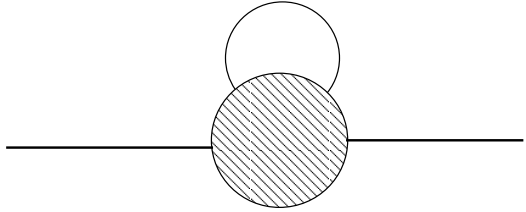,width=7cm,height=2.8cm}}
\caption{ The virial expansion of self-energy in medium to first order.}
\end{figure}

\section{Discussion}

It is well-known that the effective theory incorporating the symmetries of
the QCD Lagrangian, called the chiral perturbation theory, can successfully
describe the strong interaction processes in the low energy region. This
theory finds a natural application in the realm of hadronic statistical
physics \cite{Gasser,Gerber,Goity}.
There is also a phenomenological approach in statistical mechanics
for interacting systems, namely the method of virial expansion.

In a region where the expansions of both the methods converge rapidly, the
virial expansion would prove to be an identity in chiral perturbation
theory. However, there are situations, due to the proximity of resonances,
for example, where the effective coupling constants in the chiral Lagrangian
can be rather large and the leading term in the chiral expansion may hold only
in a limited region of interest. The virial expansion, on the other hand,
may enjoy a wider range of validity.

The case of the nucleon self-energy at finite temperature
illustrates this point \cite{Leutwyler}. Due to the presence of
$\Delta (1232)$ resonance near the $\pi N$ threshold, the chiral
expansion converges slowly, whereas
the first term in the virial expansion is a good representation for low
enough pion densities. Another example is the nucleon self-energy in
nuclear medium, which involves the interaction of the two-nucleon system.
Here the presence of bound or virtual two-nucleon states close to the
threshold  of $NN$ scattering makes it difficult to formulate a
satisfactory chiral perturbation theory for this
system \cite{Kaplan}. But the virial formula is expected to hold
for densities close to the nuclear saturation density \cite{Sandreas}.

The proofs for the virial expansion described here are indeed simple, mainly
because we restrict to the first order formula.
But the methods are not
restricted to the first term in any way. They may well provide simple
alternative proofs in the real time formalism for the complete virial
expansion \cite{Jeon}. One has only to take care of two additional aspects,
namely, the disconnected parts
that result from opening more than one internal lines and the $2\otimes 2$
matrix structure due to the so-called ghost vertices.

\section*{Acknowlegements}

I wish to thank J. Gasser and H. Leutwyler for helpful discussions.
I also thank the members of the Institute of Theretical Physics,
University of Berne, Switzerland for their kind hospitality. I acknowledge
the support of CSIR, Government of India.

\end{document}